\documentclass[aps,preprint,amssymb,12pt,floatfix]{revtex4}
\usepackage{subfigure}
\usepackage{graphicx}
\usepackage{lscape,graphicx}
\usepackage{rotating}
\usepackage{epstopdf}
\usepackage{color}
\usepackage{amsmath,amsthm}

\begin{document}
\title{Exploring the Energy Landscape of Biopolymers Using Single Molecule Force Spectroscopy and Molecular Simulations}
\author{Changbong Hyeon}
\thanks{Email: hyeoncb@kias.re.kr}
\affiliation{School of Computational Sciences, Korea Institute for Advanced Study, Seoul 130-722, Republic of Korea}
\date{\today}
\begin{abstract}
In recent years, single molecule force techniques have opened a new avenue to decipher the folding landscapes of biopolymers by allowing us to watch and manipulate the dynamics of individual proteins and nucleic acids. 
In single molecule force experiments, quantitative analyses of measurements employing sound theoretical models and molecular simulations play central role more than any other field.  
With a brief description of basic theories for force mechanics and molecular simulation technique using self-organized polymer (SOP) model, this chapter will discuss various issues in single molecule force spectroscopy (SMFS) experiments, which include pulling speed dependent unfolding pathway, measurement of energy landscape roughness, the influence of molecular handles in optical tweezers on measurement and molecular motion, and folding dynamics of biopolymers under force quench condition. 
\end{abstract}
\maketitle

The advent of single molecule (SM) techniques over the past decades has brought a significant impact on the studies of biological systems \cite{Moy94Science,Ha96PNAS,Bustamante97Science}. 
The spatial and temporal resolutions and a good force control attained in SM techniques have been used to decipher the microscopic basis of self-assembly processes in biology.   
Among SM techniques, single molecule force spectroscopy (SMFS) have been adapted not only to stretch biopolymers \cite{Bustamante94SCI,Seol07PRL} but also to unravel the internal structures and functions of many proteins and nucleic acids \cite{GaubSCI97,GaubJMB99,Bustamante03Science,Mickler07PNAS,Greenleaf08Science}. 
By precisely restricting the initial and final conformations onto specific regions of energy landscape, SMFS also provided a way to probe the collapse or folding dynamics under mechanical control, which fundamentally differs from those under temperature or denaturant control \cite{Fernandez04Science,HyeonMorrison09PNAS}. 
The observables that are usually inaccessible to conventional bulk experiments, for instance the heterogeneity of dynamic trajectories and intermediate state ensembles, have been measured to provide glimpses to the topography of complex folding landscapes \cite{Russell02PNAS,Mickler07PNAS}.  
Furthermore, the use of SMFS is being expanded to study the function of biological motors \cite{Coppin96PNAS,Visscher99Nature,Block03Science,BlockPNAS06,Chemla05Cell,Bustamante09Science,Guydosh09Nature} and cells \cite{Sheetz06NRMCB}. 

Given that foldings of biopolymers are realized through a number of elementary processes,   
good controls over time, length, force and energy scales are essential to resolve the details of biomolecular self-assembly  \cite{Greenleaf05PRL,Bustamante01Sci,Meiners10PRL}. 
The ability to control the energy scale within the range of $\sim k_BT$, in particular,  ($k_BT\approx$4.1 $pN\cdot nm$ at room temperature $T=300$ K) allows us to study how biological systems, that are evolved to accommodate the thermal fluctuations, versatilely adapt their conformation to a varying environment.   
SMFS is an excellent tool to decompose the energy required to disrupt non-covalent bonds, responsible for the stability of biological structures ($\sim\mathcal{O}(1)$ $k_BT$), into $\sim$pN force and $\sim$nm length scale.  
A phenomenological interpretation of bond rupture due to an external force in the context of cell-cell adhesion process and 
theoretical estimate of mechanical force associated with the process had already been discussed as early as in 1978 by Bell \cite{Bell78SCI}. However, only after 1990s with SMFS its experimental realization was achieved \cite{Tees93BJ,Florin94Science,Lee94Langmuir,Hoh92JACS,Alon95Nature}.   

Many biological processes, in vivo, are in fact mechanically controlled. 
The ability to apply pN-force to a single molecule and watch its motion at nm scale (or vice versa) has a great significance in molecular biology in that one can elucidate the microscopic and structural origin of a biological process by quantify both kinetic and thermodynamic properties of the biomolecule at single molecule level and compare them with those from ensemble measurements \cite{TinocoARBBS04}.
Under a constant force condition using force-clamp method \cite{Bustamante01Sci,Block03PNAS,Fernandez04Science,Visscher99Nature} near at the transition region, the molecular extension ($x$) exhibits discrete jumps among basins of attractions as a function of time. 
This immediately allows to study the hopping kinetics between the basins of attraction. If the time traces are long enough to sample all the conformations then one can also construct an equilibrium free energy profile under tension $f$ \cite{Woodside06PNAS,Block06Science}. 
The fraction of nativeness $\phi_N(f)$ or equilibrium constant $K_{eq}(f)=(1-\phi_N(f))/\phi_N(f)$ as a function of $f$ can be accurately measured just like the one using calorimetry or denaturant titration in bulk experiments. 
 
Since the first single molecule force experiment, interplay between theory, simulation and experiment have affected the experimental design as well as theoretical formulation to interpret results from measurements.   
A need to understand biomolecular dynamics at SM level further highlights the importance of theoretical background such as polymer physics \cite{deGennesbook}, stochastic theory \cite{Kramers40Physica,Hanggi90RMP} and fluid dynamics. 
Molecular simulations of SMFS using a simple model provide a number of microscopic insights that cannot be easily gained through experiments alone.     
This chapter encompasses the force mechanics from the perspectives of theories and molecular simulations. 
Basic theories for force mechanics and the main simulation technique using self-organized polymer (SOP) model will be described, followed by a number of findings and predictions for SMFS made through the concerted efforts using force theories and molecular simulations.

\section*{Force Extension Curve}
Mechanical response of a molecule is expressed with two conjugate variables, force ($\vec{f}$) and molecular extension ($\vec{x}$), to define a mechanical work ($W=\vec{f}\cdot\vec{x}$). 
Force-extension curves (FECs) or the time dependences of $f(t)$ and $x(t)$ are the lowest level  data that can infer all the relevant information concerning the mechanical response of the biomolecules. 
For a generic homopolymer whose Flory radius is $R_F\sim N^{\nu}a$, where $N$ is the number of monomers and $a$ is the size of monomer, 
the extension of polymer $x$ should be determined by a comparison between $R_F$ and tensile screening length $\xi_p=k_BT/f$. 
The force value $f$ determines the parameter $q=R_F/\xi_p$, which satisfies $q\ll 1$ ($q\gg 1$) for small (large) $f$. 
The applied force can roughly be classified into three regimes. 
(i) For small $f$, $x\ll R_F$ and $q\ll 1$ are satisfied.
Thus, $x\approx \beta R_F^2 f$ is obtained from a scaling argument $x=R_F\Phi(q)\approx  R_F q$ since $\Phi(q)\sim q$ for $q\ll 1$. 
(ii) For an intermediate $f$ ($R_F< x \ll Na$), the shape of globular polymer is distorted to form a string of tensile blobs, where blob size is given $\xi_p\sim N_b^{\nu}$ with $N_b$ being the number of monomers consisting the blob. 
The total extension of the string of tensile blobs under tension is $x\approx \xi_p\times (N/N_b)$, leading to Pincus scaling law $x\sim f^{1/\nu-1}\sim f^{2/3}$ \cite{Pincus76Macromol}. 
Here, note that this scaling law is only observed when $1\ll (\xi_p/a)^{1/\nu}\ll N$ is ensured \cite{Morrison07Macro}. 
(iii) For extremely large forces, chain is fully stretched; $x\approx Na^2\beta f/3$ for an extensible chain and $x\approx Na$ for an inextensible chain.  
 
The scaling argument for biopolymers deviates from that of generic homopolymers with $N\gg 1$ due to the finite size effect as well as various local and nonlocal interactions \cite{Morrison07Macro}. 
In practice, the persistence length ($l_p$) and contour length ($L$) of biopolymers are extracted by employing a force-extension relation ($f=\frac{k_BT}{l_p}\left[1/4(1-x/L)^2-1/4+x/L\right]$ \cite{Marko95Macro}) of worm-like chain (WLC) model, whose energy hamiltonian takes into account the bending energy penalty along the polymer chain ($H/k_BT=\frac{l_p}{2}\int^L_0\left(\frac{\partial {\bf u}(s)}{\partial s}\right)^2ds$ where ${\bf u}(s)$ is the tangential unit vector at position  $s$ along the contour) \cite{Bustamante94SCI}. 
The rips in FEC due to the disruption of internal bonds and subsequent increase in the contour length from $L$ to $L+\Delta L$ are used to decipher the energetics and internal structure of proteins and nucleic acids.  
\cite{Visscher99Nature,Bustamante01Sci,Bustamante03Science}.  
The FEC of repeat proteins demonstrates multiple peaks with saw-tooth pattern, suggesting that under tension the repeat proteins unfolds one domain after another.  
As a more complicated system, \emph{Tetrahymena} ribozyme with nine subdomains show saw-tooth patterns but with varying peak height and position, demanding more careful and laborious tasks of analysis \cite{Bustamante03Science}. 

\section*{Forced Unfolding at Constant Force}
A phenomenological  description of the forced-unbinding of adhesive contacts by Bell \cite{Bell78SCI} has played a central role in studying the force induced dynamics of biomolecules for the last two decades. 
In the presence of external force $f$, Bell modified Eyring's transition state theory \cite{EyringJCP35} as follows : 
\begin{equation}
k=\kappa\frac{k_BT}{h}e^{-(E^{\ddagger}-\gamma f)/k_BT}
\label{eqn:Erying}
\end{equation}
where $k_B$ is the Boltzmann constant, $T$ is the temperature,
$h$ is the Planck constant, and $\kappa$ is the transmission coefficient.
The parameter  $\gamma$ is  a characteristic length of the system
associated with bond disruption.
Under tension $f$, the activation barrier $E^{\ddagger}$ responsible for a stable bond is reduced to $E^{\ddagger}-\gamma\times f$.  
The prefactor $\frac{k_BT}{h}$ is the vibrational frequency of a bond due to thermal fluctuation prior to disruption.

Although the original Bell model correctly describes the stochastic nature of bond disruption, the prefactor $k_BT/h$ fails to capture the physical nature of attempt frequency for the ligand unbinding from catalytic site or the protein unfolding dynamics under tension, which depends on the shape of potential as well as viscosity of media. 
In fact, Eyring's transition state theory is only applicable for the chemical reaction in gas phase. 
More appropriate theory for the dynamics in condensed media should account for the effect of solvent viscosity and conformational diffusion \cite{Hyeon05BC}.  
One-dimensional reaction coordinate, projected from a multidimensional energy landscape, can well represent the dynamics provided that the relaxation times of conformational dynamics along a reaction coordinate is much slower than other degrees of freedom \cite{ZwanzigBook}.  
Under tension $f$, molecular extension ($x$) (or end-to-end distance ($R$)) are assumed to be a good reaction coordinate. 
On the one-dimensional reaction coordinate, mean first passage time obeys the following simple differential equation \cite{Hyeon07JP}.  
$\mathcal{L}^{\dagger}_{FP}(x)\tau(x)=-1$
where  $\mathcal{L}^{\dagger}_{FP}\equiv e^{F_{eff}(x)/k_BT}\partial_xD(x)e^{-F_{eff}(x)/k_BT}\partial_x$ is the adjoint Fokker Planck operator.
The mean first passage time of a quasi-particle between the interval $a\leq x\leq b$ with reflecting $\partial_x\tau(a)=0$ and absorbing boundary condition $\tau(b)=0$ is 
\begin{equation}
\tau(x)=\int^b_x dye^{F_{eff}(y)/k_BT}\frac{1}{D(y)}\int^y_a dz e^{-F_{eff}(z)/k_BT}.
\label{eqn:mfpt}
\end{equation}
Above, the free energy profile $F(x)$ is considered being ``tilted" by an external force by $f\cdot x$. 
As long as the transition barrier ($\Delta F^{\ddagger}=F(x_{ts})-F(x_b)$) is large enough,  
the Taylor expansions of the free energy potential $F(x)-fx$ at the barrier top and the bound state position with a saddle point approximation result in the seminal Bell-Kramers equation \cite{Kramers40Physica,Hanggi90RMP},
\begin{eqnarray}
k(f)\approx\frac{\omega_b\omega_{ts}}{2\pi\gamma}e^{\beta(\Delta F^{\ddagger}-f\Delta x^{\ddagger})}
\label{eqn:Kramers}
\end{eqnarray}
where $\Delta x^{\ddagger}\equiv x_{ts}-x_b$, $\omega_b$ and  $\omega_{ts}$ are the curvatures of the potential, $|\partial^2_xF(x)|$, at $x=x_b$ and $x_{ts}$, respectively,
and $\gamma=k_BT/Dm$ is a friction coefficient associated with the motion of biomolecule.
Experimentally determined speed limit of the folding dynamics (barrierless folding time) of two-state proteins is $\approx (0.1-1)$ $\mu s$ \cite{Hyeon05BC,GruebeleNature03,Chung09PNAS}. 
A care should be taken not to use the prefactor $(k_BT/h)^{-1}\approx 0.2$ $ps$ from transition state theory for gas phase when estimating the barrier height from folding or unfolding kinetics data of biopolymers in condensed phase.     
A cautionary word is in place. 
If the barrier height $\Delta F^{\ddagger}-f \Delta x^{\ddagger}$ is comparable or smaller than $k_BT$, the molecular configuration trapped as a metastable state in the free energy barrier can move almost freely across the barrier. 
In face, the barrier vanishes when $f$ reaches a critical force $f_c=\Delta F^{\ddagger}/\Delta x^{\ddagger}$. 
In this case, there is no separation of time scales between diffusive motion and barrier crossing event. Hence, the saddle point approximation taken for Eq.\ref{eqn:Kramers} from Eq.\ref{eqn:mfpt} does not hold. 
Due to thermal noise the unfolding or rupture event of the system occurs under finite free energy barrier ($> k_BT$) before $f$ reaches $f_c$.  
For Bell-Kramers equation to be applicable $f$ should be always smaller than $f_c$.

\section*{Forced Unfolding at Constant Loading Rate - Dynamic Force Spectroscopy} 
Even though the constant force (force clamp) experiment is more straightforward for analysis, 
due to technical reasons many of the force experiments have been performed under a constant loading condition (force-ramp) in which the force is linearly ramped over time \cite{Bustamante02Science,Bustamante03Science,FernandezNature99,FernandezTIBS99}.  
Dynamic force spectroscopy (DFS) probes the energy landscape of biomolecular complexes by detecting the mechanical response of the molecules. 
The linearly increasing mechanical force with a rate of $r_f = df/dt$ is exerted on the molecular system until the molecular complex disrupts. 
Upon unbinding, the force recorded on the instrument drops abruptly, thus one can measure the unbinding force of the system of interest (Fig,1A). 
Because of stochastic nature of unbinding event, the unbinding force of molecular complex
is not unique, but distributes broadly, defining the unbinding force distribution ($P(f)$) (Fig.1B).

Under linearly varying force ($f=r_f\times t$), the rate of barrier crossing from bound to unbound state (or from folded to unfolded state) is also time-dependent.
Hence, the survival probability at time $t$ is given by 
$S(t)=\exp{\left(-\int^t_0d\tau k(\tau)\right)}$. Thus, 
the first passage time distribution is $P(t)=-dS(t)/dt=k(t)S(t)$. 
Change of variable from $t$ to $f$ leads to a unbinding force distribution 
\begin{equation}
P(f)=\frac{1}{r_f}k(f)S(f)=\frac{1}{r_f}k(f)\exp{\left[-\int^f_0df'\frac{1}{r_f}k(f')\right]}. 
\label{eqn:force_distribution}
\end{equation}
Note that $k(f)$ is exponentially increasing function of $f$ while $S(f)$ is exponentially decreasing function of $f$ with greater power at $f\gg 1$, shaping a Gumbel distribution, $P(f)\sim e^fe^{-e^f}$, for $k(f)\sim e^f$.    
Current theoretical issue of deciphering the underlying energy landscape using force hinges on an analysis of $P(f)$ by building not only a physically reasonable but also a mathematically tractable model. 

The most probable unfolding force is obtained using $dP(f)/df|_{f=f^*}=0$. 
\begin{equation} f^*=\frac{k_BT}{\Delta x^{\ddagger}}\log{r_f}+\frac{k_BT}{\Delta x^{\ddagger}}\log{\left(\frac{\Delta x^{\ddagger}}{\nu_De^{-\beta\Delta F^{\ddagger}}k_BT}\right)}
\label{eqn:most_force}
\end{equation}
where $\nu_D\equiv \omega_o\omega_{ts}/2\pi\gamma$. 
In conventional DFS theory, Eq.\ref{eqn:most_force} is employed to extract $\Delta x^{\ddagger}$ and $\Delta F^{\ddagger}$ of underlying 1-D free energy profile associated with force dynamics. 
For unbinding force $f^*$ to be compatible to the one in the picture of Kramers barrier crossing dynamics,  
$f^*<f_c$ should be obeyed as mentioned above. 
The condition $f^*-f_c=\frac{k_BT}{\Delta x^{\ddagger}}\log{\frac{r_f\Delta x^{\ddagger}}{\nu_Dk_BT}}<0$ demands $r_f<r_f^c\left(=\frac{\nu_Dk_BT}{\Delta x^{\ddagger}}\right)$.   
For a set of parameters, $k_BT\approx 4$ $pN\cdot nm$, $\Delta x^{\ddagger}\sim 1$ $nm$, and $\nu_D\sim 10^6$ $s^{-1}$, the critical loading rate is $r_f^c\sim 10^6$ $pN/s$.
The typical loading rate used in force experiments ($0.1 pN/s <r_f < 10^3 pN/s$) is several orders of magnitude smaller than this value.  
Therefore, in all likelihood unbinding dynamics in typical experimental conditions obey stochastic barrier crossing dynamics. 
In contrast, the steered molecular dynamics simulations with all atom representation \cite{SchultenBJ98} typically uses $r_f>r_f^c$ due to high computational cost. 
In such an extreme condition, however, the forced-unfolding process can no longer be considered a thermally activated barrier crossing process.  
At high $r_f>r_f^c$ it was shown that an average rupture force ($\langle f\rangle$) grows as $r_f^{1/2}$ \cite{HummerBJ03}. 

It is of particular interest that for a molecular system unfolding through a single free energy barrier,  
the force dependence of force clamp kinetics can be formally expressed with the $P(f)$ from force ramp experiment as follows \cite{Dudko08PNAS,Dudko06PRL,Dudko07BJ}.  
\begin{equation}
k(f)=\frac{P(f)/r_f}{1-\int^f_0df' P(f')/r_f}
\end{equation}
which is easily shown using the relation, $S(f)=\exp{\left(-\int^f_0df' \frac{1}{r_f}k(f')\right)=1-\int^f_0df'\frac{1}{r_f} P(f')}$. 
Technically the two distinct experimental methods are connected through this simple relationship. 
Therefore, by conducting force-ramp experiment with a sufficiently good statistics to get $P(f)$ at varying $r_f$, one can, in principle, build a data for $k(f)$ as in force-clamp experiment.

\section*{Deformation of Energy Landscape under Tension}
Basic assumption of Bell-Kramers equation is that an external force changes the free energy barrier along the reaction coordinate from $\Delta F^{\ddagger}$ to $\Delta F^{\ddagger}-f\Delta x^{\ddagger}$ without significantly changing other topology of energy landscape.   
A  linear regression  of both Eqs. \ref{eqn:Kramers} and \ref{eqn:most_force}  provides the characteristic length $\Delta x^{\ddagger}$ and free energy barrier $\Delta F^{\ddagger}$ at $f\rightarrow 0$. 
However, in practice, nonlinearity (negative curvature for $f$ vs $\log{k(f)}$, positive curvature for $\log{r_f}$ vs $f^*$) is often detected especially when $f$ or $r_f$ is varied over broad range. 
Thus, in case $f$ (or $r_f$) is varied at a large but narrow range of $f$ (or $r_f$) then substantial errors can arise in the extrapolated values of $\Delta x^{\ddagger}$ and $\Delta F^{\ddagger}$ to the zero force ; the linear regression will underestimate $\Delta x^{\ddagger}$ and overestimate $\Delta F^{\ddagger}$.  
The physical origin of $f$-dependent $\Delta x^{\ddagger}$ is found in a complicated molecular response to the external force. 
If the TS ensemble is broadly spread along the reaction coordinate then the molecule can adopt diverse structures along the energy barrier with varying $f$ values. 
Whereas, if the TS ensemble is sharply localized along the reaction coordinate, the nature of TS ensemble measured in $x$-coordinate will be insensitive to the varying $f$ values.     
Or, more simply, the origin of moving transition state position can be algebraically explained by plotting the shape of $F_{eff}(x)$ with varying $f$. 
Because $x_{ts}$ and $x_b$ are determined from the \emph{force dependent condition} $F'(x)-f=0$,
all the parameters should be intrinsically $f$-dependent as $\Delta F^{\ddagger}(f)$, $\Delta x^{\ddagger}(f)$, $\omega_{ts}(f)$, and $\omega_b(f)$. 
By making harmonic approximation of $F(x)$ at $x=x_b$ and $x=x_{ts}$, i.e., 
$F(x)\approx F(x_b)+1/2\cdot |F^{\prime\prime}(x_b)|(x-x_b)^2$ and $F(x)\approx F(x_{ts})-1/2\cdot |F^{\prime\prime}(x_{ts})|(x-x_{ts})^2$ and calculating $\Delta x^{\ddagger}(f)=x_{ts}(f)-x_{b}(f)$ from $F^{\prime}_{eff}(x)=F'(x)-f=0$,  one can show that 
\begin{equation}
\frac{\Delta x^{\ddagger}(f)}{\Delta x^{\ddagger}}=1-\chi(f)=1-\frac{f}{\Delta x^{\ddagger}}\left(\frac{1}{|F^{\prime\prime}(x_{ts})|}+\frac{1}{|F^{\prime\prime}(x_{b})|}\right).
\end{equation} 
Typically for biopolymers under tension, free energy profile near native state minimum is sharp ($|F^{\prime\prime}(x_b)|\gg 1$). 
To minimize the difference between $\Delta x^{\ddagger}(f)$ and $\Delta x^{\ddagger}$ and to make $\chi(f)\approx 0$, the transition barrier should be sharp for a given $f/\Delta x^{\ddagger}$ value  (i.e. $f/\Delta x^{\ddagger}|F^{\prime\prime}(x_{ts})|\ll 1$).   
Simulation studies \cite{Hyeon05PNAS,Lacks05BJ,Hyeon06BJ} in which  the free energy profiles  were explicitly computed from thermodynamic considerations alone clearly showed the
change of $\Delta x^{\ddagger}$ when $f$ is varied. 
Note that the movement of transition barrier location toward the native state position ($x=x_b$) is consistent with the Hammond postulate \cite{HammondJACS53,LefflerSCI53} that explicates the nature of transition state of a simple organic compound when product state is relatively more stabilized than reactant state.   

To account for the nonlinear response of biological systems to the force more naturally, Dudko and coworkers proposed to use an analytically tractable microscopic model for the underlying free energy profile. 
For a cubic potential $F(x)=-\frac{2\Delta F^{\ddagger}}{(\Delta x^{\ddagger})^3}x^2(x-\frac{3}{2}\Delta x^{\ddagger})$ whose distance to the transition state and free energy barrier are $\Delta x^{\ddagger}$ and $\Delta F^{\ddagger}$, all the parameters needed for Kramers equation are expressed as a function of $f$, $\Delta x^{\ddagger}$ and $\Delta F^{\ddagger}$;  
$\omega_{ts}(f)\omega_b(f)=\omega_{ts}\omega_b(1-f/f_c)^{1/2}$, $\Delta x^{\ddagger}(f)=\Delta x^{\ddagger}(1-f/f_c)^{1/2}$, and $\Delta F^{\ddagger}(f)=\frac{2}{3}\Delta x^{\ddagger}f_c(1-f/f_c)^{3/2}$ where $f_c\equiv 3\Delta F^{\ddagger}/2\Delta x^{\ddagger}$. Thus, the $f$-dependent unfolding rate $k(f)$ for cubic potential is given exactly with $\nu=2/3$.  
\begin{equation}
k(f)=k(0)\left(1-\nu\frac{f\Delta x^{\ddagger}}{\Delta F^{\ddagger}}\right)^{1/\nu-1}e^{-\Delta F^{\ddagger}\left\{\left(1-\nu\frac{f\Delta x^{\ddagger}}{\Delta F^{\ddagger}}\right)^{1/\nu}-1\right\}}
\label{eqn:Dudko}
\end{equation} 
In fact, with different $\nu$ value, the same expression with Eq.\ref{eqn:Dudko} is obtained for harmonic-cusp potential ($\nu=1/2$) and Bell-Kramers model ($\nu=1$).   
The precise value of $\nu$ depends on the nature of the underlying potential and could be treated as an adjustable parameter \cite{Dudko06PRL}.
Consequently, cusp, cubic \cite{Dudko03PNAS,Dudko06PRL,Dudko08PNAS} or piecewise harmonic potentials \cite{Freund09PNAS} have been suggested as microscopic models for underlying free energy profile.  
So far, theories for force experiments have been devised mainly for single barrier picture in one-dimensional reaction coordinate. 
Although two slope fit using multiple energy barrier picture was suggested to explain the large curvature observed in DFS data \cite{EvansNature99}, building an multibarrier free energy profile from one-dimensional information such as $P(f)$ or $[\log r_f, f^*]$ curve is an inverse problem whose answer may not be unique, as is well demonstrated by Derenyi et al. in the context of the forced unfolding over two sequentially located transition barriers \cite{Ajdari04BJ}. 

\section*{Self-Organized Polymer (SOP) Model for Single Molecule Force Spectroscopy}
At present, the typical spatial resolution reached in SMFS is a few nm, and the dynamics that are probed in SMFS are rather global than local. 
The details of local dynamics such as the breakage of a particular hydrogen bond or an isomerization of dihedral angle cannot be discerned from FECÕs or the time trace of molecular extension alone. 
To gain and provide sufficient insight through molecular simulations in conjunction with SMFS, a simple model with which one can efficiently simulate the forced unfolding dynamics of a large biopolymer at a spatial resolution of SMFS would be of great use. 
By drastically simplifying the details of local interactions such as bond angle or dihedral angle along the backbone, which indeed are the major determinant for the dynamics under tension at nm scale, one can either gain an acceleration in simulation speed or explore the dynamics of a larger molecule. 
As is well appreciated in the literature of normal-mode analysis, an inclusion of small length scale information do not alter the global dynamics corresponding to a low frequency mode \cite{TirionPRL96,BaharCOSB05,Zheng06PNAS}.  
Self-organized polymer model, proposed in this line of thought, is well suited to simulate the forced unfolding dynamics of large biopolymers at spatial resolution of SMFS.   
The basic idea of the SOP model is to use the simplest possible energy hamiltonian to faithfully reproduce the topology of native fold and to simulate the low-resolution global dynamics of biopolymers of arbitrary size \cite{Hyeon06Structure,HyeonBJ07,Hyeon06PNAS,Lin08JACS,Chen10PNAS,Dima08PNAS,Hyeon07PNAS2}. 
The energy function for biopolymers in the SOP representation is
\begin{align}
E_{SOP}(\{\vec{r}_i\})&=E_{FENE}+E_{nb}^{(att)}+E_{nb}^{(rep)}\nonumber\\
&=-\sum_{i=1}^{N-1}\frac{k}{2}R_0^2\log({1-\frac{(r_{i,i+1}-r_{i,i+1}^o)^2}{R_0^2}})\nonumber\\
&+\sum_{i=1}^{N-3}\sum_{j=i+3}^N\epsilon_h\left[\left(\frac{r^o_{ij}}{r_{ij}}\right)^{12}-2\left(\frac{r^o_{ij}}{r_{ij}}\right)^6\right]\Delta_{ij}+\sum_{i<j}\epsilon_l\left(\frac{\sigma}{r_{ij}}\right)^6(1-\Delta_{ij}).
\label{eqn:SOP}
\end{align}
The first term in Eq. \ref{eqn:SOP} is the finite extensible nonlinear elastic (FENE) potential for chain
connectivity with parameters, $k = 20 kcal/(mol\cdot A2)$, $R_0 = 0.2$ nm, $r_{i,i+1}$ is the distance between
neighboring beads at $i$ and $i+1$, and $r^0_{i,i+1}$ is the distance in the native structure. 
The use of the FENE potential for backbone connectivity is more advantageous than the standard harmonic potential, especially for forced- stretching to produce an inextensible behavior of WLC. 
The Lennard-Jones potential is used to account for interactions that stabilize the native topology. 
A native contact is defined for bead pairs $i$ and $j$ such that $|i-j| > 2$ and whose distance is less than 8 \AA\ in the native state. 
We use $\epsilon_h = 1-2$ $kcal/mol$ for native pairs, and $\epsilon_l = 1$ $kcal/mol$ for nonnative pairs. 
In the current version, we have neglected nonnative attractions. 
This should not qualitatively affect the results, because under tension such interactions are greatly destabilized. 
To ensure noncrossing of the chain, $(i,i+2)$ pairs interacted repulsively with $\sigma = 3.8$ \AA. 
There are five parameters in the SOP force field. In principle, the ratio of $\epsilon_h/\epsilon_l$ and $R_c$ can be adjusted to obtain realistic values of critical forces. 
For simplicity, we choose a uniform value of $\epsilon_h$ for all protein constructs. $\epsilon_h$ can be made sequence-dependent and ion-implicit as $\epsilon_h\rightarrow \epsilon_h^{ij}$ if one wants to improve the simulation results.
By truncating forces due to the Lennard-Jones potential
for interacting pairs with $r > 3r_{ij}^0$ or $3\sigma$, the computational cost essentially scales as $\sim \mathcal{O}(N)$. 
We refer to the model as the Ôself-organized polymerÕ (SOP) model because it only uses the polymeric nature of the biomolecules and the crucial topological constraints that arise from the specific fold. 

With SOP representation, the dynamics of biopolymers are simulated under force-clamp or force-ramp condition by solving the equation of motions in the overdamped regime. 
The position of the $i^{th}$ bead at time $t+h$ is given by 
\begin{equation}
x_i(t+h)=x_i(t)+\frac{h}{\zeta}\left(F_i(t)+\Gamma_i(t)+f_N(t)\right)
\end{equation} 
where $F_i(t)=-\nabla_{x_i}E_{SOP}(\{\vec{r}_i\})$, the $x$-component of the conformational force acting on the $i^{th}$ bead, and 
$\Gamma(t)$ is a random force selected from a Gaussian noise distribution $P[\Gamma_i(t)]\propto \exp{\left[-\frac{1}{4k_BT\zeta}\int^t_0d\tau\Gamma_i^2(\tau)\right]}$. 
The conversion of simulation time into the physical time is made by using $\partial_t r\approx a/\tau\sim \zeta^{-1}(k_BT/a)$ and 
$m a^2/\epsilon_h=\tau_L^2$, 
\begin{equation}
\tau_H\sim \frac{\zeta a^2}{k_BT}=\left[\frac{\zeta(\tau_L/m)\cdot \epsilon_h}{k_BT}\right]\tau_L. 
\end{equation}
When amino acid residue is used as a coarse-grained center, $\zeta \approx (50 - 100)$ $m/\tau_L$. The time step $h\approx 0.1 \tau_L$. 
For force-clamp condition, a constant force ($f_N$) is exerted to the $N^{th}$ bead $f_N(t)=f_N$ with the 1st bead being fixed, $f_1(t)=0$. 
For force-ramp condition, $f_N(t)=-k(x_N-vt)$ with $f_1(t)=0$ is used. A harmonic spring with stiffness $k$ is attached to the $N^{th}$ bead and the position of spring is moved with a constant velocity $v$. 

\section*{Deciphering the Energy Landscape of Complex Biomolecules using SMFS} 
The mechanical response of the molecule becomes more complex with an increasing complexity in the native topology of biomolecules.
This section will address two classes of unfolding scenarios for the molcules with complex topology. 

First, it is conceivable that a biomolecule with complicated topology in its native state is unravelled via more than two distinct transition state ensemble so that the forced-unfolding routes bifurcate \cite{Li08PNAS,Mickler07PNAS}.  
In the scenario, the survival probability of molecule remaining in the native state decays as 
\begin{equation}
S_N(t)=\sum_{i=1}^N\varphi_i(f)\exp{(-t/\tau_i(f))}
\end{equation} 
where unfolding time along the $i$-th route is given by $\tau_i(f)$ and $\varphi_i(f)$ is the partition factor for $i$-th route with $\sum_{i=1}^N\varphi_i(f)=1$, both of which are the function of $f$ \cite{Hyeon05BC,Zwanzig90ACR}.  
Mechanical response of a barrel shaped Green fluorescence protein (GFP), made of 11 $\beta$-strands with one $\alpha$-helix at the N-terminal, is quite intricate, whose unfolding path depends on pullling speed and direction \cite{Mickler07PNAS,Dietz06PNAS}. 
In earlier force experiment on GFP by Rief and coworkers,  the intricacy of the GFP forced-unfolding is manifested as the indistinguishability of two unfolding routes \cite{Dietz04PNAS}. 
After the $\alpha$-helix being disrupted from the barrel structure due to external force, the second rip is due to the peeling off of $\beta 1$ or $\beta 11$. The gains of contour length ($\Delta L$) from the molecular rupture event from $\beta 1$ and from $\beta$11 are, however, identical, so that it was impossible to identify the source of the second peak simply by analyzing the FEC \cite{Dietz04PNAS}.  
The force simulation using SOP model suggested a bifurcation into the two different routes by showing that 70 \% of the molecule disrupt from N-terminal and the remainder of the molecules from C-terminal.  
Experimentally, this is confirmed by fixing either N-terminal or C-terminal direction by introducing a disulfide bridge through mutation \cite{Mickler07PNAS}. 
The kinetic partitioning mechanism (KPM) used for protein and RNA folding \cite{ThirumalaiACR96,Hyeon05BC} can be adapted to explain the mechanical behavior of biopolymers. 

Second, reaction coordinate under tension can have sequentially aligned multiple barriers.    
In comparison to GFP, the forced-unfolding experiment on RNase-H was characterized by a peculiar mechanical response \cite{Marqusee05Science}.  
The FEC of RNase-H has a single large rip along unfolding path but has two rips in refolding path. 
This behavior was explained by considering a shape of free energy profile where native ($N$), intermediate ($I$), and unfolding ($U$) state lie sequentially with relatively high transition barrier ($\gg k_BT$) between $N$ and $I$, and low transition barrier ($\sim k_BT$) between $I$ and $U$.   
On such a free energy profile, the unfolding to the $U$ state would occur by skipping $I$ state because the external force that disrupt to the $N$ to overcome the first barrier is already larger than the mechanical stability of the $I$ state relative to $U$. 
Thus,  Accessing to the $I$ state is difficult from the $N$ under an increasing tension while $I$ can be reached from $U$ in the refolding FEC since the free energy barrier between $I$ and $U$ is relatively small ($\sim k_BT$). 
This hypothetical picture was further supported by the force-clamp method. For $I\rightleftharpoons U$, the transition mid-force is $f_{m,I\rightleftharpoons U}\approx 5.5$ pN while the escape from native basin of attraction (NBA) occurs at $f\approx 15-20$ pN \cite{Marqusee05Science}. 
Even for a molecule hopping between $I$ and $U$ at $f=5.5$ pN, the molecule can get to $N$ state. However, once the molecule jumps over the barrier between $N$ and $I$, being trapped in $N$. the molecule has little chance to jump back to $I$ within the measurement time.    

\section*{Measurement of Energy Landscape Roughness}
Although the energy landscape of biopolymers are evolutionary tailored such that the potential gradient toward the native state is large enough to drive the biopolymers to their native state, the energetic and topological frustration still remain to render the folding landscape rugged, slowing down the folding processes.   
To account for the effect of energy landscape roughness in a one-dimensional free energy profile, $F(x)$ can be effectively decomposed into $F(x)=F_0(x)+F_1(x)$ \cite{ZwanzigPNAS88}.
where $F_0(x)$ is a smooth potential that determines  the global shape of the energy landscape, and $F_1(x)$ is the ruggedness that superimposes $F_0(x)$.  
By taking the spatial average over $F_1(x)$  using
$\langle e^{\pm\beta F_1(x)}\rangle_l=\frac{1}{l}\int^l_0dx e^{\pm\beta F_1(x)}$, where $l$ is the ruggedness length scale, the
 associated mean first passage time is altered to 
$\tau(x)\approx\int^b_x dye^{F_0(y)/k_BT}\langle e^{\beta F_1(y)}\rangle_l\frac{1}{D}\int^y_a dz e^{-F_0(z)/k_BT}\langle e^{-\beta F_1(z)}\rangle_l$.
By either assuming a Gaussian distribution of the roughness contribution $F_1$  ($P(F_1)\propto e^{-F_1^2/2\epsilon^2}$) or simply assuming $\beta F_1\ll 1$ and $\langle F_1\rangle=0$, $\langle F_1^2\rangle=\epsilon^2$ and $\beta\epsilon$ is small, the effective diffusion coefficient can be approximated as $D^*\approx D\exp{\left(-\beta^2\epsilon^2\right)}$ where $D$ is the bare diffusion constant.

The signature of the roughness of the underlying energy landscape is uniquely reflected in the
non-Arrhenius temperature dependence of the unbinding rates. 
By using mean first passage time with the effective diffusion coefficient with roughness, one can show that the unfolding kinetics for a two-state folder deviates substantially from an Arrhenius behavior as follows.
\begin{equation}
\log{k(f,T)}=a+b/T-\varepsilon^2/T^2
\label{eqn:k_mod}
\end{equation}
where $\varepsilon^2$ is a constant even if the  coefficients $a$ and $b$ change under different force and temperature conditions \cite{Hyeon03PNAS}.  
This relationship suggests that conducting \emph{force-clamp experiments} over the range of temperatures identifies the roughness scale $\varepsilon$. 
Here the condition $\epsilon/\Delta F^{\ddagger}\ll 1$ should be ensured. 

To extract the roughness scale, $\epsilon$, using DFS, 
a series of DFS experiments should be performed as
a function of $T$ and $r_f$ so that reliable unfolding force distributions ($P(f)$) and corresponding $f^*$ value are obtained.
\begin{equation}
f^*\approx\frac{k_BT}{\Delta x^{\ddagger}}\log{r_f}+\frac{k_BT}{\Delta x^{\ddagger}}\log{\frac{\Delta x^{\ddagger}}{\nu_De^{-\Delta F_0^{\ddagger}/k_BT} k_BT}}+\frac{\varepsilon^2}{\Delta x^{\ddagger} k_BT}.
\label{eqn:most_force_approx}
\end{equation}
One way of obtaining the $\varepsilon$ from experimental data is as follows \cite{Reich05EMBOrep}.  From the $f^*$ vs $\log{r_f}$ curves at two different temperatures, $T_1$ and $T_2$,
one can obtain $r_f(T_1)$ and $r_f(T_2)$ for which the  $f^*$ values are identical.
By equating the right-hand side of the expression in Eq.\ref{eqn:most_force} at $T_1$ and $T_2$
the scale $\epsilon$ can be estimated \cite{Hyeon03PNAS,Reich05EMBOrep} as
\begin{small}
\begin{align}
\varepsilon^2 &\approx\frac{\Delta x^{\ddagger}(T_1)k_BT_1\times\Delta x^{\ddagger}(T_2)k_BT_2}{\Delta x^{\ddagger}(T_1)k_BT_1-\Delta x^{\ddagger}(T_2)k_BT_2}\nonumber\\
&\times \left[\Delta F^{\ddagger}_0\left(\frac{1}{\Delta x^{\ddagger}(T_1)}-\frac{1}{\Delta x^{\ddagger}(T_2)}\right)+\frac{k_BT_1}{\Delta x^{\ddagger}(T_1)}\log{\frac{r_f(T_1)\Delta x^{\ddagger}(T_1)}{\nu_D(T_1)k_BT_1}}-\frac{k_BT_2}{\Delta x^{\ddagger}(T_2)}\log{\frac{r_f(T_2)\Delta x^{\ddagger}(T_2)}{\nu_D(T_2)k_BT_2}}\right]. 
\label{eqn:epsilon}
\end{align}
\end{small}
This equation has been used to measure $\varepsilon$ for GTPase Ran$-$Importin $\beta$ complex ($\varepsilon>5k_BT$) \cite{Reich05EMBOrep} and transmembrane helices ($\varepsilon \approx 4-6$ $k_BT$) \cite{Janovjak07JACS}.

\section*{Pulling Speed Dependent Unfolding Pathway}
Dynamics of polymer is extremely intricate due to multiply entangled length and time scales.
Polymeric nature of RNA and proteins immediately lends itself when a molecule interacts with an external force. 
A biopolymer adapts its configuration in response to an external stress in a finite amount of relaxation time.  
In both AFM and LOT experiments, a force is applied to one end of the chain with the other end being fixed. 
A finite amount of delay is expected for the tension $f$ to propagate along the backbone of molecule and through the network of contacts that stabilize the native topology. 
To understand the effect of finite propagation time of the tension on unfolding dynamics, 
it is useful to consider a ratio between the loading rate $r_f$ and the rate at which the applied force propagates along polymer chain $r_T$ ($\lambda=r_T/r_f$). 
$r_f$ is controlled by experiments; $r_T$ most likely depends on the topology of a molecule. 
Depending the value of parameter $\lambda$, the history of dynamics can be altered qualitatively. 
If $\lambda\gg  1$, then the applied tension at one end propagates rapidly so that, even prior to the realization of the first rip, force along the chain is uniform. 
In the opposite limit, $\lambda\ll  1$, the tension is nonuniformly distributed along the backbone at the moment any of rupture event occurs (see the gradient of red color in the one at the highest loading rate, the top panel of Fig.2A). 
In such a situation, unraveling of RNA begins from a region where the value of local force exceeds the tertiary interactions. 

The unfolding simulation using SOP model of \emph{Azoarcus} ribozyme provides a great insight into the issue of force propagation and $r_f$-dependent unfolding pathways \cite{Hyeon06Structure}. 
The intuitive argument given in Fig.2A is clarified by visualizing the change in the pattern of force propagation for \emph{Azoarcus} ribozyme under three different loading conditions.  
Alignment of the angles between the bond segment vector ($r_{i,i+1}$) along force direction can tell the magnitude of force exerted at each position along the backbone.  
The nonuniformity in the local segmental alignment is most evident at the highest loading rate. 
The dynamics of the force propagation occurs sequentially from one end of the chain to the other at high $r_f$. 
The alignment of segment along $f$ gets more homogeneous at lower $r_f$. 
These results highlight an important prediction closely related to polymer dynamics, that the unfolding pathways can drastically change depending on the loading rate, $r_f$. 
By varying $\lambda$ (i.e., controlling $r_f$) from $\lambda\ll 1$ to $\lambda \gg 1$, force experiments will show the dramatic effect of pulling speed dependence on the unfolding dynamics. 
There may be a dramatic change in unfolding mechanism for two different instruments using dinstinct different $r_f$ (LOT and AFM experiments). 
In addition, predictions of mechanism for forced unfolding based on all-atom MD simulations \cite{Schulten09Structure} should also be treated with caution unless due to topological reason, the unfolding pathways are robust  to large variations in the loading rates regardless of $\lambda$ value.

\section*{Effect of Molecular Handles on the Measurement of Hopping Transition Dynamics}
While the idea of SM experiment is to probe the dynamics of an isolated molecule, noise or interference from many possible sources is always a difficult problem to deal with in nano-scale measurement.  
To accurately measure the dynamics of a test molecule the interference between molecule and instrument should be minimal. 
In optical tweezers experiment, dsDNA or DNA-RNA hybrid handles are inserted between the microbead and test molecule so as to minimize the systematic error due to the microbead-molecule interaction. 
Unlike force-ramp, issues concerning non-equilibrium relaxation such as the delay of signal or force propagation does not lend themselves in force clamp experiments as long as the time scale of molecular hopping ($\tau_{hop}$) at given force is greater than the relaxation times of handle $(\tau_h)$, microbead $(\tau_b)$, and other part of the instruments (typically $\tau_{hop}\gg \tau_h, \tau_b$). Force-clamp experiment essentially creates an equilibrium condition. As a result, tension is uniformly distributed over the handle as well as the test molecule while the molecule hops.    

However, critical issue as to measurements still remains even at a perfect equilibrium due to \emph{handle fluctuations}. 
(i) Since the dynamics of test molecule is measured by monitoring the position of microbead, 
to gain an identical signal with the test molecule, the fluctuation of the handle should be minimal.   
Therefore, one may conclude that \emph{short and stiff} handles (smaller $L/l_p$) are ideal for precisely sampling the conformational statistics \cite{Block06Science} (see Fig.3B and caption). 
(ii) Simulations of hopping dynamics with handles differing length and flexibility shows that the hopping kinetics is least compromised from the true handle-free kinetics when \emph{short and flexible} handles are used \cite{Manosas07BJ}.  
Physically, when a test molecule is sandwiched between handles, the diffusive motion is dynamically pinned. 
The stiffer handles, the slower the hopping transition. Therefore, \emph{short and flexible} handles are suitable to preserve the true dynamics of the test molecule. 

The above two conditions for (i) precise measurement of thermodynamics and (ii) accurate measurement of true kinetics apparently contradict each other.     
However, this dilemma can be avoided by simply measuring the folding landscape accurately using \emph{short and stiff} handle as long as the effective diffusion coefficient associated with the reaction coordinate is known. 
At least, at $f\approx f_{mid}$, the hopping time trace can easily fulfill the ergodic condition, providing a good equilibrium free energy profile. Furthermore, it can be shown that the molecular extension $z$ (or the end-to-end distance $R$) used to represent the free energy profile is indeed a good reaction coordinate on which the Bell's prescription to obtain rate agrees well with the folding rate one can obtain using a simulation;  for a given transition rate from NBA to UBA $k_F(f_{mid})$, the variation of force from $f=f_{mid}$ modifies the rate as $k_F(f_{mid}\pm \delta f)=k_F(f_{mid})\exp{(\pm \delta f\cdot \Delta x^{\ddagger}/k_BT)}$.    
Once an accurate free energy ($F(z_m)\approx F^o(z_m)\approx F(z_{sys})$ holds for $L/l_p\ll 1$. see Fig.3B and caption) is obtained, one can determine the hopping kinetics by directly calculating the mean first passage time on $F^o(z_m)$. 
The only unknown, the effective diffusion coefficient on $F^o(z_m)$, can be estimated by mapping the hopping dynamics of test molecule to the dynamics of a simple analytical model such as generalized Rouse model (GRM) \cite{Barsegov08PRL,Hyeon08PNAS}.

\section*{Folding Dynamics upon Force-Quench}
Because of the multi-dimensional nature, the dynamics of biomolecules is sensitive to the condition to which the molecule is imposed. 
In a rugged energy landscape, the folding rate is not unique. Rather, the folding kinetics can vary greatly depending on initial conditions \cite{Russell02PNAS}.  
The difference in the initial condition may be due to either urea concentration, temperature, or force. 
For RNA it is well known that the initial counterion condition can alter the folding route in a drastic fashion \cite{Russell02PNAS}.  
A biopolymer of interest adapts its structure to a given condition and creates a ``condition specific denatured state ensemble (DSE)."
The routes to the NBA from the DSE are determined by the shape and ruggedness of energy landscape due to the factors such as side-chain interactions, topological frustrations and so forth.  

The force quench refolding dynamics of poly-ubiquitin (poly-Ub) by Fernandez and Li \cite{Fernandez04Science} is the first experiment that focused on the folding dynamics of proteins from a fully extended ensemble, in which the dynamics of molecular extension of poly-Ub construct was traced from an initial stretched ensemble prepared at high stretch-force $f_S=122$ pN to a quenched ensemble at low quench-force $f_Q=15$ pN (see Fig.4A). 
The folding trajectories monitored using the molecular extension ($R$) was at least an order of magnitude slower than the one from bulk measurement and was characterized with continuous transitions divided into at least three (four in \cite{Fernandez04Science}) stages: initial reduction of $R$, long flat plateau, and a cooperative collapse transition at the final stage.
Two main questions were immediately raised from the results of experiments. 
(i) Why is the folding (collapse) process so slow compared to the one at bulk measurement? 
(ii)  What is the nature of the plateau and cooperative transition at the final stage? 
Concerning the point (i), it was suspected that an aggregate was formed between the Ub monomer since the effective concentration is more than the critical concentration for the aggregate formation \cite{Sosnick04SCI}. 
But, this possibility was ruled out since no signature of disrupting the aggregate contacts was found when the folded poly-Ub construct was re-stretched; the number of step indicating the unfolding of individual Ub domain was consistent with the number of Ub monomers
\cite{Fernandez04Science}. 

The immediate interpretation toward the anomalous behavior of refolding trajectories upon force quench is found from the vastly different initial structural ensemble \cite{FernandezTIBS99,Li06PNAS,Hyeon06BJ,Hyeon08JACS,HyeonMorrison09PNAS}. 
The initial structural ensemble under high tension is fully stretched (SSE, stretched state ensemble) while the nature of ensemble for bulk measurements are thermally denatured (thermally denatured ensemble, TDE). 
In terms of structural characteristics, TDE and SSE have drastic difference. 
The entropy of SSE is smaller than TDE ($S_{SSE}<S_{TDE}$) \cite{HyeonMorrison09PNAS,Hyeon08JACS,Hyeon06BJ}.
Therefore, it is not unusual that the folding kinetics upon force quench is vastly different from the one at the bulk measurement. 
It is conceivable that the energy landscape explored starting from these two distinct ensembles vastly differ.  

The next more elaborate question is then why the folding rate for force-quench is slower. 
Under force quench condition, two driving forces compete each other. One is the free energy gradient bias toward the NBA, the other is the quench force ($f_Q<f_{mid}$) that resists the collapse process. A free energy barrier is formed under these two competing forces. Consequently, prior to making a transition to the NBA, the molecule is trapped in a finite sized free energy barrier, forming a metastable intermediate. Due to $f_Q$, the formation of contact responsible for the collapse process is suppressed. This tendency increases with an increasing $f_Q$, which again increases the refolding time ($\tau_F$). If the time spent for force quench ($\tau_Q$) is too fast, i.e., $\tau_Q\ll \tau_F$, producing a nonequilibrated system in which the molecule is trapped in the force-induced metastable intermediate (FIMI), then the plateau in terms of molecular extension $x$ will be observed in the measurement. 
Compared to TDE where the contacts responsible for the collapse are in proximity, the formation of folding (or collapse) nuclei is much more time-consuming for a system trapped in FIMI state at higher $f_Q$.  
Thus, the folding route from SSE to NBA differs significantly from the one from TDE to NBA \cite{Hyeon08JACS}.
 
The FIMI is generic to the force-quench refolding dynamics of any biopolymer, which can be easily tested by applying a tension to a semiflexible polymer that forms a toroid or racquets structures upon collapse. 
The energy hamiltonian is given by
\begin{equation}
	  \mathcal{H}=\frac{k_s}{2a^2}\sum_{i=1}^{N-1}(r_{i,i+1}-a)^2+\frac{k_b}{2}\sum_{i=1}^{N-2}(1-\hat{r}_i\cdot\hat{r}_{i+1})+\epsilon_{LJ}\sum_{i,j}\left[\left(\frac{a}{r_{ij}}\right)^{12}-2\left(\frac{a}{r_{ij}}\right)^6\right]-f(z_{N}-z_1) \label{WLCHam.eq}
\end{equation}
with the parameters, $\epsilon_{LJ}=1.5k_B T$, $k_s=2000k_B T$, $N=200$, $a=0.6$ and $k_b=80$ $k_BT$. 
By abruptly changing $f=f_S=83$ pN to $f=f_Q=4$ pN, long plateaux with varying durations are observed in the semiflexible polymer described by the above energy hamiltonian (Fig.4B).   
An optical tweezers experiment showing the difficulty of forming DNA toroid under tension unambiguously supports the argument that FIMI is generic to collapse dynamics of any polymer under tension \cite{Fu06JACS} (Fig.4C). 

\section*{Concluding Remarks}
The effort to decipher the energy landscape of a biopolymer by monitoring its mechanical response to an external stress has greatly enriched single molecule experiments, theory and molecular simulations associated with force mechanics of biomolecules and cells.
We have also witnessed many instances of successful application of theories and molecular simulations for the experimental data analysis \cite{BarsegovPNAS05,Hyeon03PNAS,Reich05EMBOrep,Dudko07BJ}.    
However, as experiments are conducted on more complicated and larger systems among cellular constituents, non-trivial patterns of molecular response are more observed \cite{EvansNature99,Chu09PNAS,NevoNSB03,Zhu03Nature}.  
Theoretical method to analyze the mechanical response from multiple barriers and studies on the structural origin of more non-trivial mechanical responses such as catch-slip bond are still on their infancy.   
Further studies need to be done beyond force induced dynamics of a molecular system associated with a single free energy barrier crossing dynamics on a one-dimensional energy profile \cite{Hyeon07JP,BarsegovPNAS05,Suzuki10PRL}.  
Interference between a molecule of interest and instrument itself pose the problem of deconvolution \cite{Hyeon08PNAS,Block06Science}. 
Of great interest from the perspective of both theory and simulation will be revealing how hydrodynamic interaction between the subdomains of a biopolymer with complex architecture play a role during its forced-unfolding process. 
More exciting experiments using mechanical force and breakthroughs in theories and molecular simulation methods are anticipated in the next decade to further reveal the beauty of living systems at the microscopic level. 
\\

I am grateful to Dave Thirumalai for intellectually pleasurable discussions and collaboration on theory for force spectroscopy and molecular simulations for the last many years. This work was supported in part by grants from National Research Foundation of Korea (R01-2008-000-10920-0).


\begin{thebibliography}{10}

\bibitem{Moy94Science}
Moy, V.~T, Florin, E.~L,  \& Gaub, H.~E.
\newblock (1994) Intermolecular forces and energies between ligands and
  receptors.
\newblock {\em Science} {\bf 266}, 257--259.

\bibitem{Ha96PNAS}
Ha, T, Enderle, T, Ogletree, D.~F, Chemla, D.~S, Selvin, P,  \& Weiss, S.
\newblock (1996) {Probing the interaction between two single molecules:
  Fluorescence resonance energy transfer between a single donor and a single
  acceptor}.
\newblock {\em Proc. Natl. Acad. Sci. USA} {\bf 93}, 6264.

\bibitem{Bustamante97Science}
Kellermeyer, M.~Z, Smith, S.~B, Granzier, H.~L,  \& Bustamante, C.
\newblock (1997) {Folding-Unfolding Transitions in Single Titin Molecules
  Characterized by Force-Measuring Laser Tweezers}.
\newblock {\em Science} {\bf 276}, 1112--1116.

\bibitem{Bustamante94SCI}
Bustamante, C, Marko, J.~F, Siggia, E.~D,  \& Smith, S.
\newblock (1994) Entropic elasticity of $\lambda$-phase {DNA}.
\newblock {\em Science} {\bf 265}, 1599--1600.

\bibitem{Seol07PRL}
Seol, Y, Skinner, G.~M, Visscher, K, Buhot, A,  \& Halperin, A.
\newblock (2007) {Stretching of Homopolymeric RNA Reveals Single-Stranded
  Helices and Base-Stacking}.
\newblock {\em Phys. Rev. Lett.} {\bf 98}, 158103.

\bibitem{Bustamante03Science}
Onoa, B, Dumont, S, Liphardt, J, Smith, S.~B, {Tinoco, Jr.}, I,  \& Bustamante,
  C.
\newblock (2003) {Identifying Kinetic Barriers to Mechanical Unfolding of the
  \emph{T. thermophila} Ribozyme}.
\newblock {\em Science} {\bf 299}, 1892--1895.

\bibitem{Mickler07PNAS}
Mickler, M, Dima, R.~I, Dietz, H, Hyeon, C, Thirumalai, D,  \& Rief, M.
\newblock (2007) {Revealing the bifurcation in the unfolding pathways of GFP by
  using single-molecule experiments and simulations}.
\newblock {\em Proc. Natl. Acad. Sci. USA} {\bf 104}, 20268--20273.

\bibitem{GaubSCI97}
Rief, M, Gautel, H, Oesterhelt, F, Fernandez, J.~M,  \& Gaub, H.~E.
\newblock (1997) {Reversible Unfolding of Individual Titin Immunoglobulin
  Domains by AFM}.
\newblock {\em Science} {\bf 276}, 1109--1111.

\bibitem{GaubJMB99}
Rief, M, Pascual, J, Saraste, M,  \& Gaub, H.~E.
\newblock (1999) Single molecule force spectroscopy of spectrin repeats: low
  unfolding forces in helix bundles.
\newblock {\em J. Mol. Biol.} {\bf 286}, 553--561.

\bibitem{Greenleaf08Science}
Greenleaf, W.~J, Frieda, K.~L, Foster, D. A.~N, Woodside, M.~T,  \& Block,
  S.~M.
\newblock (2008) Direct observation of hierarchical folding in single
  riboswitch aptamers.
\newblock {\em Science} {\bf 319}, 630--633.

\bibitem{Fernandez04Science}
Fernandez, J.~M \& Li, H.
\newblock (2004) Force-clamp spectroscopy monitors the folding trajectory of a
  single protein.
\newblock {\em Science} {\bf 303}, 1674--1678.

\bibitem{HyeonMorrison09PNAS}
Hyeon, C, Morrison, G, Pincus, D.~L,  \& Thirumalai, D.
\newblock (2009) Refolding dynamics of stretched biopolymers upon force-quench.
\newblock {\em Proc. Natl. Acad. Sci. USA} {\bf 106}, 20288--20293.

\bibitem{Russell02PNAS}
Russell, R, Zhuang, X, Babcock, H, Millett, I, Doniach, S, Chu, S,  \&
  Herschlag, D.
\newblock (2002) Exploring the folding landscape of a structured {RNA}.
\newblock {\em Proc. Natl. Acad. Sci. USA} {\bf 99}, 155--160.

\bibitem{Visscher99Nature}
Visscher, K, Schnitzer, M.~J,  \& Block, S.~M.
\newblock (1999) Single kinesin molecules studied with a molecular force clamp.
\newblock {\em Nature} {\bf 400}, 184--187.

\bibitem{BlockPNAS06}
Guydosh, N.~R \& Block, S.~M.
\newblock (2006) Backsteps induced by nucleotide analogs suggest the front head
  of kinesin is gated by strain.
\newblock {\em Proc. Natl. Acad. Sci. USA} {\bf 103}, 8054--8059.

\bibitem{Chemla05Cell}
Chemla, Y.~R, Aathavan, K, Michaelis, J, Grimes, S, Jardine, P.~J, Anderson,
  D.~L,  \& Bustamante, C.
\newblock (2005) {Mechanism of Force Generation of a Viral DNA Packaging
  Motor}.
\newblock {\em Cell} {\bf 122}, 683--692.

\bibitem{Bustamante09Science}
Hodges, C, Bintu, L, Lubkowska, L, Kashlev, M,  \& Bustamante, C.
\newblock (2009) {Nucleosomal Fluctuations Govern the Transcription Dynamics of
  RNA Polymerase II}.
\newblock {\em Science} {\bf 325}, 626--628.

\bibitem{Coppin96PNAS}
Coppin, C.~M, Finer, J.~T, Spudich, J.~A,  \& Vale, R.~D.
\newblock (1996) Detection of sub-8-nm movements of kinesin by high-resolution
  optical-trap microscopy.
\newblock {\em Proc. Natl. Acad. Sci. USA} {\bf 93}, 1913--1917.

\bibitem{Guydosh09Nature}
Guydosh, N.~R \& Block, S.~M.
\newblock (2009) Direct observation of the binding state of the kinesin head to
  the microtubule.
\newblock {\em Nature} {\bf 461}, 125--128.

\bibitem{Block03Science}
Asbury, C.~L, Fehr, A.~N,  \& Block, S.~M.
\newblock (2003) {Kinesin Moves by an Asymmetric Hand-Over-Hand Mechanism}.
\newblock {\em Science} {\bf 302}, 2130--2134.

\bibitem{Sheetz06NRMCB}
Vogel, V \& Sheetz, M.
\newblock (2006) Local force and geometry sensing regulate cell functions.
\newblock {\em Nature Rev. Mol. Cell Biol.} {\bf 7}, 265--275.

\bibitem{Greenleaf05PRL}
Greenleaf, W.~J, Woodside, M.~T, Abbondanzieri, E.~A,  \& Block, S.~M.
\newblock (2005) Passive all-optical force clamp for high-resolution laser
  trapping.
\newblock {\em Phys. Rev. Lett.} {\bf 95}, 208102.

\bibitem{Bustamante01Sci}
Liphardt, J, Onoa, B, Smith, S.~B, {Tinoco Jr.}, I,  \& Bustamante, C.
\newblock (2001) {Reversible Unfolding of Single RNA Molecules by Mechanical
  Force}.
\newblock {\em Science} {\bf 292}, 733--737.

\bibitem{Meiners10PRL}
Chen, Y.~F, Milstein, J.~N,  \& Meiners, J.~C.
\newblock (2010) {Protein-Mediated DNA Loop Formation and Breakdown in a
  Fluctuating Environment}.
\newblock {\em Phys. Rev. Lett.} {\bf 104}, 258103.

\bibitem{Bell78SCI}
Bell, G.~I.
\newblock (1978) Models for the specific adhesion of cells to cells.
\newblock {\em Science} {\bf 200}, 618--627.

\bibitem{Tees93BJ}
Tees, D. F.~J, Coenen, O,  \& Goldsmith, H.~L.
\newblock (1993) {Interaction forces between red cells agglutinated by
  antibody. IV. Time and force dependence of break-up}.
\newblock {\em Biophys. J.} {\bf 65}, 1318--1334.

\bibitem{Florin94Science}
Florin, E.~L, Moy, V.~T,  \& Gaub, H.~E.
\newblock (1994) Adhesive forces between individual ligand-receptor pairs.
\newblock {\em Science} {\bf 264}, 415--417.

\bibitem{Lee94Langmuir}
Lee, G.~U, Kidwell, D.~A,  \& Colton, R.~J.
\newblock (1994) Sensing discrete streptavidin-biotin interactions with atomic
  force microscopy.
\newblock {\em Langmuir} {\bf 10}, 354--357.

\bibitem{Hoh92JACS}
Hoh, J.~H, Cleveland, J.~P, Prater, C.~B, Revel, J.~P,  \& Hansma, P.~K.
\newblock (1992) Quantized adhesion detected with the atomic force microscope.
\newblock {\em J. Am. Chem. Soc.} {\bf 114}, 4917--4918.

\bibitem{Alon95Nature}
Alon, R, Hammer, D.~A,  \& Springer, T.~A.
\newblock (1995) Lifetime of the p-selectin-carbohydrate bond and its response
  to tensile force in hydrodynamic flow.
\newblock {\em Nature} {\bf 374}, 539--542.

\bibitem{TinocoARBBS04}
{Tinoco Jr.}, I.
\newblock (2004) {Force as a useful variable in reactions: Unfolding RNA}.
\newblock {\em Ann. Rev. Biophys. Biomol. Struct.} {\bf 33}, 363--385.

\bibitem{Block03PNAS}
Block, S.~M, Asbury, C.~L, Shaevitz, J.~W,  \& Lang, M.~J.
\newblock (2003) {Probing the kinesin reaction cycle with a 2D optical force
  clamp}.
\newblock {\em Proc. Natl. Acad. Sci. USA} {\bf 100}, 2351--2356.

\bibitem{Woodside06PNAS}
Woodside, M.~T, Behnke-Parks, W.~M, Larizadeh, K, Travers, K, Herschlag, D,  \&
  Block, S.~M.
\newblock (2006) {Nanomechanical measurements of the sequence-dependent folding
  landscapes of single nucleic acid hairpins}.
\newblock {\em Proc. Natl. Acad. Sci. USA} {\bf 103}, 6190--6195.

\bibitem{Block06Science}
Woodside, M.~T, Anthony, P.~C, Behnke-Parks, W.~M, Larizadeh, K, Herschlag, D,
  \& Block, S.~M.
\newblock (2006) Direct measurement of the full, sequence-dependent folding
  landscape of a nucleic acid.
\newblock {\em Science} {\bf 314}, 1001--1004.

\bibitem{deGennesbook}
{de Gennes}, P.~G.
\newblock (1979) {\em {Scaling Concepts in Polymer Physics}}.
\newblock (Cornell University Press, Ithaca and London).

\bibitem{Kramers40Physica}
Kramers, H.~A.
\newblock (1940) Brownian motion a field of force and the diffusion model of
  chemical reaction.
\newblock {\em Physica} {\bf 7}, 284--304.

\bibitem{Hanggi90RMP}
Hanggi, P, Talkner, P,  \& Borkovec, M.
\newblock (1990) {Reaction-rate theory: fifty years after Kramers}.
\newblock {\em Rev. Mod. Phys.} {\bf 62}, 251--341.

\bibitem{Pincus76Macromol}
Pincus, P.
\newblock (1976) Excluded volume effects and stretched polymer chains.
\newblock {\em Macromolecules} {\bf 9}, 386--388.

\bibitem{Morrison07Macro}
Morrison, G, Hyeon, C, Toan, N.~M, Ha, B.~Y,  \& Thirumalai, D.
\newblock (2007) Stretching homopolymers.
\newblock {\em Macromolecules} {\bf 40}, 7343--7353.

\bibitem{Marko95Macro}
Marko, J.~F \& Siggia, E.~D.
\newblock (1995) {Stretching DNA}.
\newblock {\em Macromolecules} {\bf 28}, 8759--8770.

\bibitem{EyringJCP35}
Eyring, H.
\newblock (1935) The activated complex in chemical reactions.
\newblock {\em J. Chem. Phys.} {\bf 3}, 107--115.

\bibitem{Hyeon05BC}
Thirumalai, D \& Hyeon, C.
\newblock (2005) {RNA and Protein folding: Common Themes and Variations}.
\newblock {\em Biochemistry} {\bf 44}, 4957--4970.

\bibitem{ZwanzigBook}
Zwanzig, R.
\newblock (2001) {\em Nonequilibrium Statistical Mechanics}.
\newblock (Oxford University press, New York).

\bibitem{Hyeon07JP}
Hyeon, C \& Thirumalai, D.
\newblock (2007) Measuring the energy landscape roughness and the transition
  state location of biomolecules using single molecule mechanical unfolding
  experiments.
\newblock {\em J. Phys.: Condens. Matter} {\bf 19}, 113101.

\bibitem{GruebeleNature03}
Yang, W.~Y \& Gruebele, M.
\newblock (2003) Folding at the speed limit.
\newblock {\em Nature} {\bf 423}, 193 -- 197.

\bibitem{Chung09PNAS}
Chung, H.~S, Louis, J.~M,  \& Eaton, W.~A.
\newblock (2009) Experimental determination of upper bound for transition path
  times in protein folding from single-molecule photon-by-photon trajectories.
\newblock {\em Proc. Natl. Acad. Sci. USA} {\bf 106}, 11839--11844.

\bibitem{Bustamante02Science}
Liphardt, J, Dumont, S, Smith, S.~B, {Tinoco, Jr.}, I,  \& Bustamante, C.
\newblock (2002) {Equilibrium information from nonequilibrium measurements in
  an experimental test of Jarzynski's equality}.
\newblock {\em Science} {\bf 296}, 1832--1835.

\bibitem{FernandezNature99}
Marszalek, P.~E, Lu, H, Li, H, Carrion-Vazquez, M, Oberhauser, A.~F, Schulten,
  K,  \& Fernandez, J.~M.
\newblock (1999) Mechanical unfolding intermediates in titin modules.
\newblock {\em Nature} {\bf 402}, 100--103.

\bibitem{FernandezTIBS99}
Fisher, T.~E, Oberhauser, A.~F, Carrion-Vazquez, M, Marszalek, P.~E,  \&
  Fernandez, J.~M.
\newblock (1999) The study of protein mechanics with the atomic force
  microscope.
\newblock {\em TIBS} {\bf 24}, 379--384.

\bibitem{SchultenBJ98}
Lu, H, Isralewitz, B, Krammer, A, Vogel, V,  \& Schulten, K.
\newblock (1998) {Unfolding of Titin Immunoglobulin Domains by Steered
  Molecular Dynamics}.
\newblock {\em Biophys. J.} {\bf 75}, 662--671.

\bibitem{HummerBJ03}
Hummer, G \& Szabo, A.
\newblock (2003) {Kinetics from Nonequilibrium Single-Molecule Pulling
  Experiments}.
\newblock {\em Biophys. J.} {\bf 85}, 5--15.

\bibitem{Dudko08PNAS}
Dudko, O.~K, Hummer, G,  \& Szabo, A.
\newblock (2008) {Theory, Analysis, and Interpretation of Single-Molecule Force
  Spectroscopy Experiments}.
\newblock {\em Proc. Natl. Acad. Sci. USA} {\bf 105}, 15755--15760.

\bibitem{Dudko06PRL}
Dudko, O.~K, Hummer, G,  \& Szabo, A.
\newblock (2006) Intrinsic rates and activation free energies from
  single-molecule pulling experiments.
\newblock {\em Phys. Rev. Lett.} {\bf 96}, 108101.

\bibitem{Dudko07BJ}
Dudko, O.~K, Math{\'e}, J, Szabo, A, Meller, A,  \& Hummer, G.
\newblock (2007) {Extracting Kinetics from Single-Molecule Force Spectroscopy:
  Nanopore Unzipping of DNA Hairpins}.
\newblock {\em Biophys. J.} {\bf 92}, 4188--4195.

\bibitem{Hyeon05PNAS}
Hyeon, C \& Thirumalai, D.
\newblock (2005) {Mechanical unfolding of RNA hairpins}.
\newblock {\em Proc. Natl. Acad. Sci. USA} {\bf 102}, 6789--6794.

\bibitem{Lacks05BJ}
Lacks, D.~J.
\newblock (2005) {Energy Landscape Distortions and the Mechanical Unfolding of
  Proteins}.
\newblock {\em Biophys. J.} {\bf 88}, 3494--3501.

\bibitem{Hyeon06BJ}
Hyeon, C \& Thirumalai, D.
\newblock (2006) Forced-unfolding and force-quench refolding of {RNA} hairpins.
\newblock {\em Biophys. J.} {\bf 90}, 3410--3427.

\bibitem{HammondJACS53}
Hammond, G.~S.
\newblock (1953) A correlation of reaction rates.
\newblock {\em J. Am. Chem. Soc.} {\bf 77}, 334--338.

\bibitem{LefflerSCI53}
Leffler, J.~E.
\newblock (1953) Parameters for the description of transition states.
\newblock {\em Science} {\bf 117}, 340--341.

\bibitem{Dudko03PNAS}
Dudko, O.~K, Filippov, A.~E, Klafter, J,  \& Urbakh, M.
\newblock (2003) Beyond the conventional description of dynamic force
  spectroscopy of adhesion bonds.
\newblock {\em Proc. Natl. Acad. Sci. USA} {\bf 100}, 11378--11381.

\bibitem{Freund09PNAS}
Freund, L.~B.
\newblock (2009) Characterizing the resistence generated by a molecular bond as
  it is forcibly separated.
\newblock {\em Proc. Natl. Acad. Sci. USA} {\bf 106}, {8818?8823}.

\bibitem{EvansNature99}
Merkel, R, Nassoy, P, Leung, A, Ritchie, K,  \& Evans, E.
\newblock (1999) Energy landscapes of receptor-ligand bonds explored with
  dynamic force spectroscopy.
\newblock {\em Nature} {\bf 397}, 50--53.

\bibitem{Ajdari04BJ}
Derenyi, I, Bartolo, D,  \& Ajdari, A.
\newblock (2004) Effects of intermediate bound states in dynamic force
  spectroscopy.
\newblock {\em Biophys. J.} {\bf 86}, 1263--1269.

\bibitem{TirionPRL96}
Tirion, M.~M.
\newblock (1996) Large amplitude elastic motions in proteins from a
  single-parameter, atomic analysis.
\newblock {\em Phys. Rev. Lett.} {\bf 77}, 1905--1908.

\bibitem{BaharCOSB05}
Bahar, I \& Rader, A.~J.
\newblock (2005) Coarse-grained normal mode analysis in structural biology.
\newblock {\em Curr. Opin. Struct. Biol.} {\bf 15}, 586--592.

\bibitem{Zheng06PNAS}
Zheng, W, Brooks, B.~R,  \& Thirumalai, D.
\newblock (2006) Low-frequency normal modes that describe allosteric
  transitions in biological nanomachines are robust to sequence variations.
\newblock {\em Proc. Natl. Acad. Sci. USA} {\bf 103}, 7664--7669.

\bibitem{Hyeon06Structure}
Hyeon, C, Dima, R.~I,  \& Thirumalai, D.
\newblock (2006) {Pathways and kinetic barriers in mechanical unfolding and
  refolding of RNA and proteins}.
\newblock {\em Structure} {\bf 14}, 1633--1645.

\bibitem{HyeonBJ07}
Hyeon, C \& Thirumalai, D.
\newblock (2007) {Mechanical unfolding of RNA : from hairpins to structures
  with internal multiloops}.
\newblock {\em Biophys. J.} {\bf 92}, 731--743.

\bibitem{Hyeon06PNAS}
Hyeon, C, Lorimer, G.~H,  \& Thirumalai, D.
\newblock (2006) {Dynamics of allosteric transition in GroEL}.
\newblock {\em Proc. Natl. Acad. Sci. USA} {\bf 103}, 18939--18944.

\bibitem{Lin08JACS}
Lin, J \& Thirumalai, D.
\newblock (2008) {Relative Stability of Helices Determines the Folding
  Landscape of Adenine Riboswitch Aptamers}.
\newblock {\em J. Am. Chem. Soc.} {\bf 130}, 14080--14081.

\bibitem{Chen10PNAS}
Chen, J, Darst, S.~A,  \& Thirumalai, D.
\newblock (2010) Promoter melting triggered by bacterial rna polymerase occurs
  in three steps.
\newblock {\em Proc. Natl. Acad. Sci. USA} {\bf 107}, 12523--12528.

\bibitem{Dima08PNAS}
Dima, R.~I \& Joshi, H.
\newblock (2008) Probing the origin of tubulin rigidity with molecular
  simulations.
\newblock {\em Proc. Natl. Acad. Sci. USA} {\bf 105}, 15743--15748.

\bibitem{Hyeon07PNAS2}
Hyeon, C \& Onuchic, J.~N.
\newblock (2007) Mechanical control of the directional stepping dynamics of the
  kinesin motor.
\newblock {\em Proc. Natl. Acad. Sci. USA} {\bf 104}, 17382--17387.

\bibitem{Li08PNAS}
Peng, Q \& Li, H.
\newblock (2008) {Atomic force microscopy reveals parallel mechanical unfolding
  pathways of T4 lysozyme: Evidence for a kinetic partitioning mechanism}.
\newblock {\em Proc. Natl. Acad. Sci. USA} {\bf 105}, 1885--1890.

\bibitem{Zwanzig90ACR}
Zwanzig, R.
\newblock (1990) Rate processes with dynamical disorder.
\newblock {\em Acc. Chem. Res.} {\bf 23}, 148--152.

\bibitem{Dietz06PNAS}
Dietz, H, Berkemeier, F, Bertz, M,  \& Rief, M.
\newblock (2006) Anisotropic deformation response of single protein molecules.
\newblock {\em Proc. Natl. Acad. Sci. USA} {\bf 103}, 12724--12728.

\bibitem{Dietz04PNAS}
Dietz, H \& Rief, M.
\newblock (2004) Exploring the energy landscape of {GFP} by single-molecule
  mechanical experiments.
\newblock {\em Proc. Natl. Acad. Sci. USA} {\bf 101}, 16192--16197.

\bibitem{ThirumalaiACR96}
Thirumalai, D \& Woodson, S.~A.
\newblock (1996) {Kinetics of Folding of Proteins and RNA}.
\newblock {\em Acc. Chem. Res.} {\bf 29}, 433--439.

\bibitem{Marqusee05Science}
Cecconi, C, Shank, E.~A, Bustamante, C,  \& Marqusee, S.
\newblock (2005) {Direct Observation of Three-State Folding of a Single Protein
  Molecule}.
\newblock {\em Science} {\bf 309}, 2057--2060.

\bibitem{ZwanzigPNAS88}
Zwanzig, R.
\newblock (1988) Diffusion in rough potential.
\newblock {\em Proc. Natl. Acad. Sci. USA} {\bf 85}, 2029--2030.

\bibitem{Hyeon03PNAS}
Hyeon, C \& Thirumalai, D.
\newblock (2003) Can energy landscape roughness of proteins and {RNA} be
  measured by using mechanical unfolding experiments?
\newblock {\em Proc. Natl. Acad. Sci. USA} {\bf 100}, 10249--10253.

\bibitem{Reich05EMBOrep}
Nevo, R, Brumfeld, V, Kapon, R, Hinterdorfer, P,  \& Reich, Z.
\newblock (2005) Direct measurement of protein energy landscape roughness.
\newblock {\em EMBO reports} {\bf 6}, 482.

\bibitem{Janovjak07JACS}
Janovjak, H, Knaus, H,  \& Muller, D.~J.
\newblock (2007) Transmembrane helices have rough energy surfaces.
\newblock {\em J. Am. Chem. Soc.} {\bf 129}, 246--247.

\bibitem{Schulten09Structure}
Lee, E.~H, Hsin, J, Sotomayor, M, Comellas, G,  \& Schulten, K.
\newblock (2009) {Discovery Through the Computational Microscope}.
\newblock {\em Structure} {\bf 17}, 1295--1306.

\bibitem{Manosas07BJ}
Manosas, M, Wen, J.~D, Li, P. T.~X, Smith, S.~B, Bustamante, C, {Tinoco, Jr.},
  I,  \& Ritort, F.
\newblock (2007) {Force Unfolding Kinetics of RNA using Optical Tweezers. II.
  Modeling Experiments}.
\newblock {\em Biophys. J.} {\bf 92}, 3010--3021.

\bibitem{Hyeon08PNAS}
Hyeon, C, Morrison, G,  \& Thirumalai, D.
\newblock (2008) {Force dependent hopping rates of RNA hairpins can be
  estimated from accurate measurement of the folding landscapes}.
\newblock {\em Proc. Natl. Acad. Sci. USA} {\bf 105}, 9604--9606.

\bibitem{Barsegov08PRL}
Barsegov, V, Morrison, G,  \& Thirumalai, D.
\newblock (2008) Role of internal chain dynamics on the rupture kinetic of
  adhesive contacts.
\newblock {\em Phys. Rev. Lett.} {\bf 100}, 248102.

\bibitem{Sosnick04SCI}
Sosnick, T.~R.
\newblock (2004) {Comment on``Force-Clamp Spectroscopy Monitors the Folding
  Trajectory of a Single Protein"}.
\newblock {\em Science} {\bf 306}, 411b.

\bibitem{Li06PNAS}
Li, M.~S, Hu, C.~K, Klimov, D.~K,  \& Thirumalai, D.
\newblock (2006) {Multiple stepwise refolding of immunoglobulin I27 upon force
  quench depends on initial conditions}.
\newblock {\em Proc. Natl. Acad. Sci. USA} {\bf 103}, 93--98.

\bibitem{Hyeon08JACS}
Hyeon, C \& Thirumalai, D.
\newblock (2008) {Multiple probes are required to explore and control the
  rugged energy landscape of RNA hairpins}.
\newblock {\em J. Am. Chem. Soc.} {\bf 130}, 1538--1539.

\bibitem{Fu06JACS}
Fu, W.~B, Wang, X.~L, Zhang, X.~H, Ran, S.~Y, Yan, J,  \& Li, M.
\newblock (2006) {Compaction Dynamics of Single DNA molecule under tension}.
\newblock {\em J. Am. Chem. Soc.} {\bf 128}, 15040--15041.

\bibitem{BarsegovPNAS05}
Barsegov, V \& Thirumalai, D.
\newblock (2005) Dynamics of unbinding of cell adhesion molecules: Transition
  from catch to slip bonds.
\newblock {\em Proc. Natl. Acad. Sci. USA} {\bf 102}, 1835--1839.

\bibitem{Zhu03Nature}
Marshall, B.~T, Long, M, Piper, J.~W, Yago, T, {McEver}, R.~P,  \& Zhu, C.
\newblock (2003) Direct observation of catch bonds involving cell-adhesion
  molecules.
\newblock {\em Nature} {\bf 423}, 190--193.

\bibitem{Chu09PNAS}
Zhang, Y, Sivasankar, S, Nelson, W.~J,  \& Chu, S.
\newblock (2009) Resolving cadherin interactions and binding cooperativity at
  the single-molecule level.
\newblock {\em Proc. Natl. Acad. Sci. USA} {\bf 106}, 109--114.

\bibitem{NevoNSB03}
Nevo, R, Stroh, C, Kienberger, F, Kaftan, D, Brumfeld, V, Elbaum, M, Reich, Z,
  \& Hinterdorfer, P.
\newblock (2003) {A molecular switch between alternative conformational states
  in the complex of Ran and importin $\beta$ 1}.
\newblock {\em Nature. Struct. Biol.} {\bf 10}, 553--557.

\bibitem{Suzuki10PRL}
Suzuki, Y \& Dudko, O.~K.
\newblock (2010) {Single-Molecule Rupture Dynamics on Multidimensional
  Landscapes}.
\newblock {\em Phys. Rev. Lett.} {\bf 104}, 048101.

\end{thebibliography}
\clearpage 
\section*{Figure Captions}

{\bf Fig.1} Force-ramp unfolding simulation of an RNA hairpin using SOP model. 
 (A) FEC ($f(R)$) at $r_f=45pN/s$ ($k=0.07pN/nm$, $v=0.64\mu m/s$). 
 $f(R)$ can be decomposed using time $t$ into $R(t)$ and $f(t)$. 
 The experimental setup of optical tweezer is mimicked by attaching a harmonic spring with strength $k$. The force is recorded by measuring the extension of this spring. 
 $f(t)$ shows how the force is ramped with time when the system is pulled with constant pulling speed $v$. The two-state hopping transition in $R$ and $f$ begins when $f$ reaches $\approx 12$ pN and ends at $f\approx 17$ pN. 
 (B) Unfolding force distribution from 100 trajectories for each $r_f$. $r_f=r_f^o$, $r_f^o/3$, $20r_f^o/3$, $10r_f^o$ when $r_f^o=4.5\times 10^3$ pN/s. 
 (C) $[\log{r_f},f^*]$ plot obtained from (B), which clearly shows a positive curvature. Linear regression at different $r_f$ values provide different $\Delta x^{\ddagger}$ values. Figure adapted from \cite{HyeonBJ07}. 
 \\
 
{\bf Fig.2} Force propagation. (A) A diagram depicting the pulling speed dependent unfolding pathway. 
 For a biopolymer consisting o two hairpins 1 and 2 whose barriers associated with forced unfolding are given $\Delta F^{\ddagger}_1$ and $\Delta F^{\ddagger}_2$ with $\Delta F^{\ddagger}_1<\Delta F^{\ddagger}_2$, the unfolding will occur through one of the two reaction paths depending on the $r_f$. 
 For $\lambda\gg 1$, tension will uniformly distribute along the chain, so that $f_1\approx f_2$. 
 Since $\Delta F^{\ddagger}_1-f_1\Delta x^{\ddagger}<\Delta F^{\ddagger}_2-f_2\Delta x^{\ddagger}$, the hairpin 1 will unfold before hairpin 2 (pathway I). 
 In contrast, for $\lambda\ll1$, $f_1\ll f_2$ leads to $\Delta F^{\ddagger}_1-f_1\Delta x^{\ddagger}>\Delta F^{\ddagger}_2-f_2\Delta x^{\ddagger}$, thus the hairpin 2 will unfold before hairpin 1 (pathway II).
 (B) $r_f$-dependent unfolding pathway simulated with \emph{Azoarcus} ribozyme using SOP model.   
FEC (middle panel) shows that unfolding occurs via $N\rightarrow [P5]\rightarrow [P6]\rightarrow [P2]\rightarrow [P4]\rightarrow [P3]\rightarrow [P1]$ at $r_f=1.2\times 10^6$ pN/s (red). $N\rightarrow [P1,P5,P6]\rightarrow [P2]\rightarrow [P4]\rightarrow [P3]$ at $r_f= 3.6\times 10^5$ pN/s (green), and $N\rightarrow [P1,P2,P5,P6]\rightarrow [P3,P4]$ at $r_f=1.8\times 10^4$ pN/s (blue). From high to low $r_f$ the unfolding pathways were changed completely. 
Time evolutions of $\cos{\theta_i}$ $(i = 1, 2, \ldots, N-1)$ at three $r_f$ visualize how the tension is being propagated.
Figure adapted from \cite{Hyeon06Structure}
\\

{\bf Fig.3} Handle effect on the measurement and RNA hopping dynamics. 
(A) Molecular simulation of RNA hopping dynamics with two handles of $L=25$ nm and $l_p=70$ nm attached to the 5' and 3' ends. 
The illustration was created by using the simulated structures collected every 0.5 ms. An example of the time trace of each component of the system, at $f=15.4$ pN. $z_m$ and $z_{sys}$ measure the extension dynamics of the RNA hairpin and of the handle-RNA-handle system, respectively. 
The time-averaged value $\overline{z}_r(t)=(1/\tau)\int^t_0d\tau z_{sys}(\tau)$ for the time trace of $z_{sys}$ is shown as the bold line. The histograms of the extension are shown on top of each column.
(B) The free energy profiles, $F_{eq}(z_{sys})$ from the two ends of handles attached (dashed line in blue),  $F_{eq}(z_{m})$ from 5' and 3' ends of RNA (solid line in red), and $F_{eq}^o(z_m)$ from 5' and 3' ends of RNA with no handles attached. When $L/l_p\lll 1$, 
$F_{eq}(z_{sys})\approx F_{eq}(z_m)\approx F^o_{eq}(z_m)$.  
(C) With increasing handle length the folding rate ($k_F^m(L)$ deviates from a true folding rate ($k_F^m(0)$). This effect is larger for handles with a greater $l_p$. 
Figure adapted from \cite{Hyeon08PNAS}. \\

{\bf Fig.4} Refolding (Collapse) dynamics of biopolymers upon force quench. 
(A) Refolding trajectory of poly-Ub generated by atomic force microscopy upon force quench (Figure adapted from \cite{Fernandez04Science}). 
(B) The collapse of a semiflexible chain in a poor solvent under force quench condition. Toroidal (magenta), racquet structures (green) are formed after varying durations of plateaux associated with FIMIs (Figure adapted from \cite{HyeonMorrison09PNAS}).
(C) After an addition of multivalent counterions, it takes as long as $\approx 2400$ sec for a $\lambda$-DNA to initiate collapse dynamics under a small quench force $f_Q=1.2$ pN (Figure adapted from \cite{Fu06JACS}). \\
\clearpage 

\begin{figure}
 \includegraphics[width=6.0in]{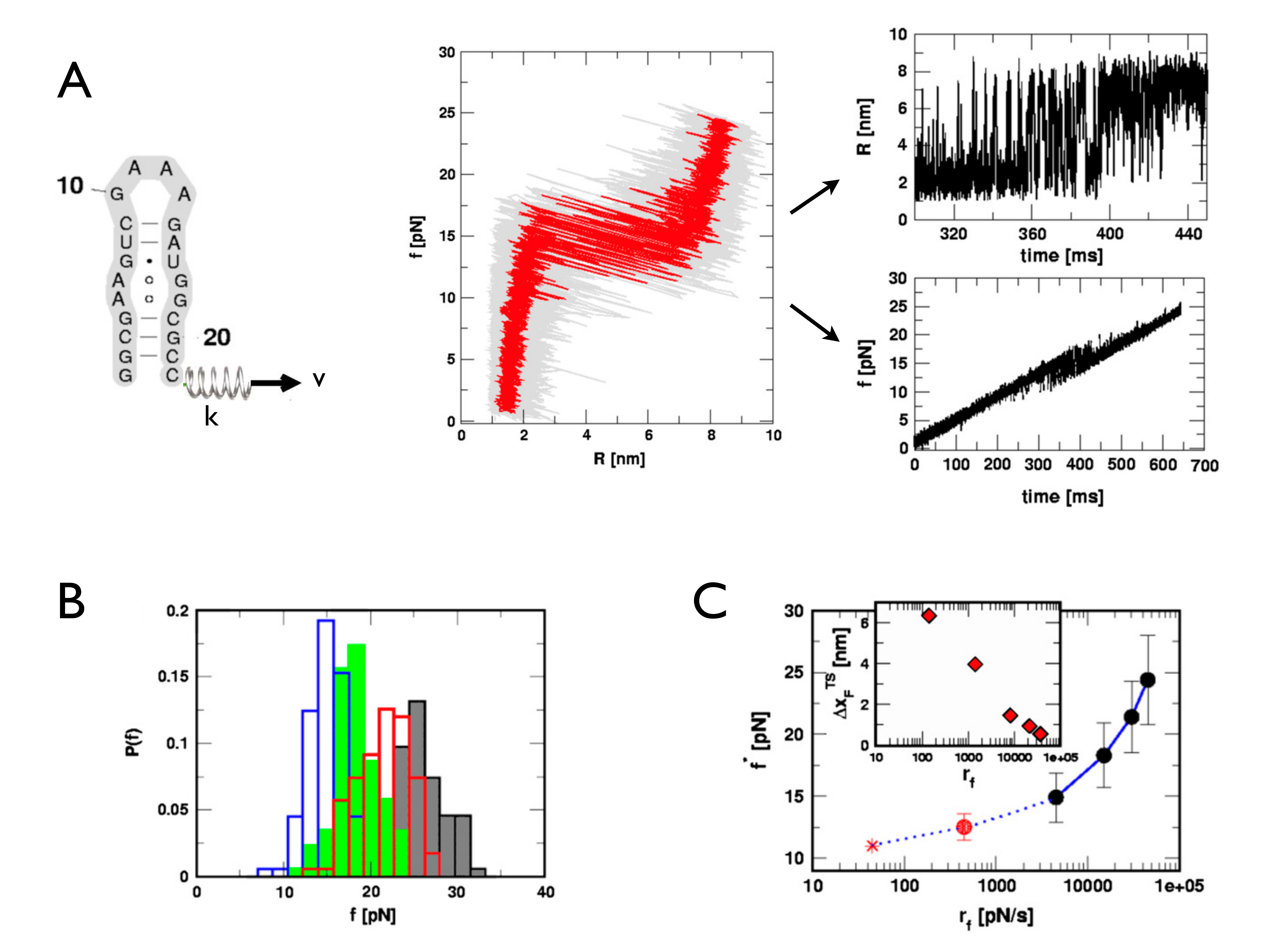}
 \caption{}
\end{figure}

\begin{figure}
 \includegraphics[width=6.0in]{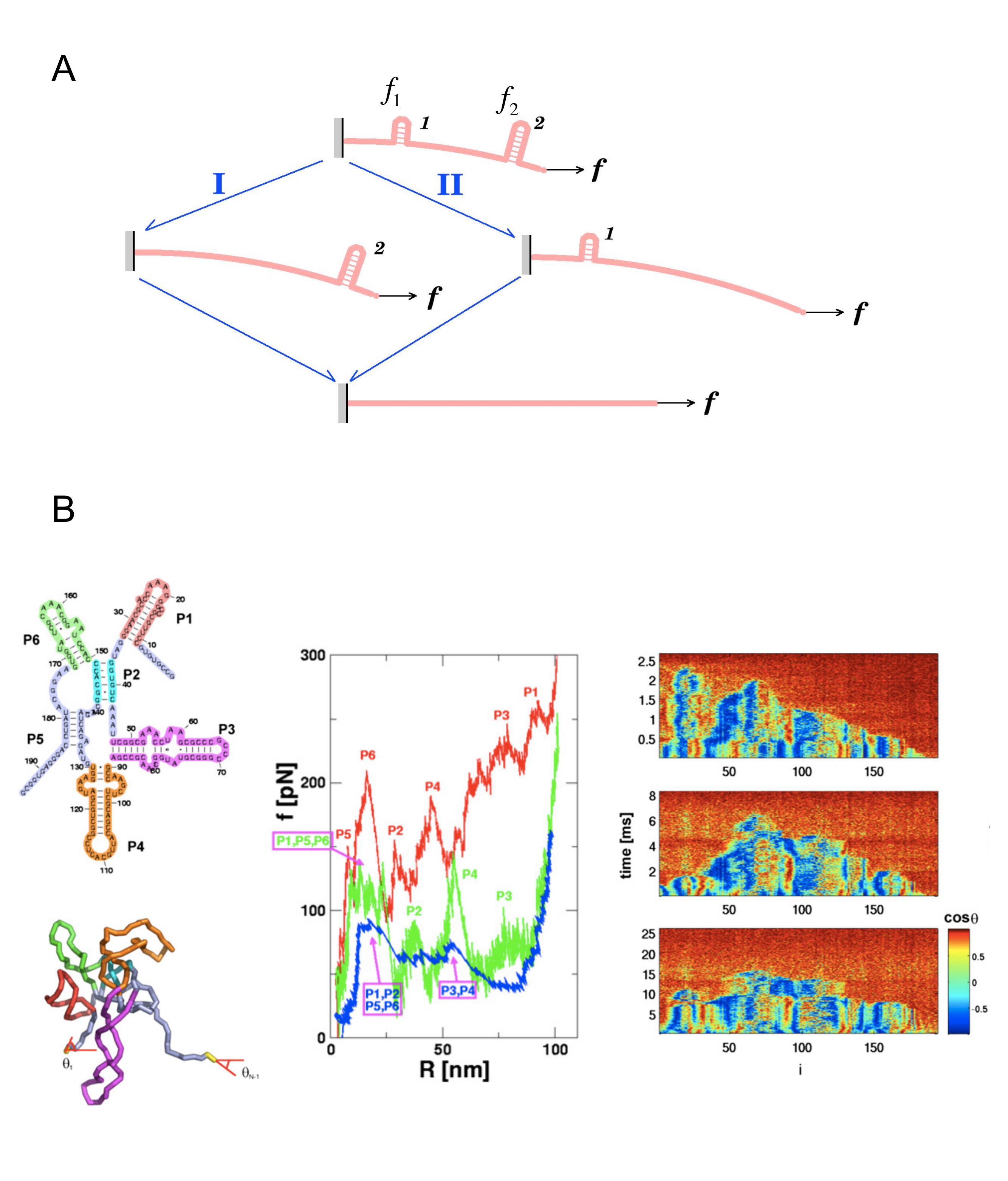}
 \caption{ }
\end{figure}

\begin{figure}
 \includegraphics[width=6.0in]{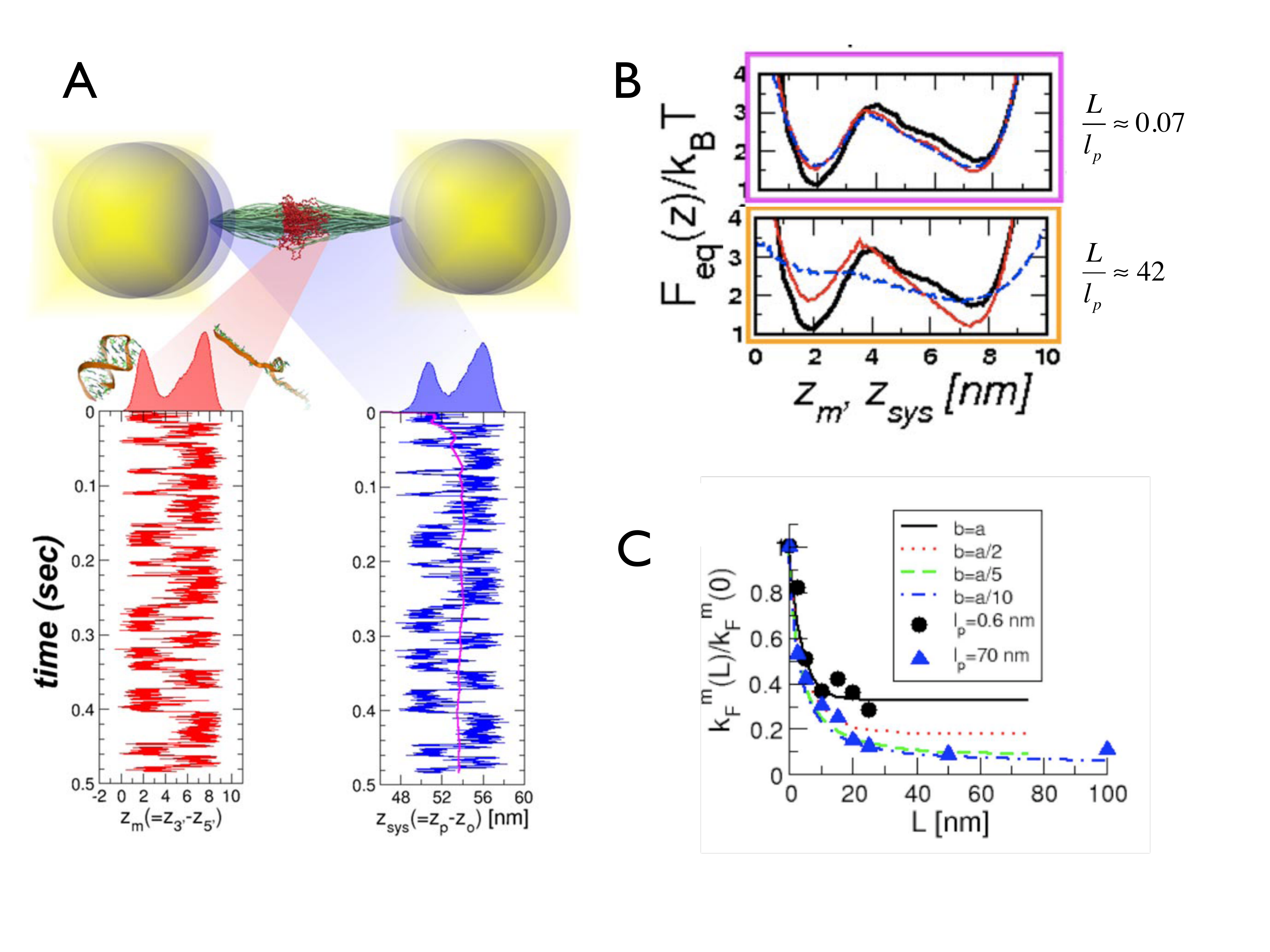}
 \caption{ }
\end{figure}

\begin{figure}
 \includegraphics[width=6.0in]{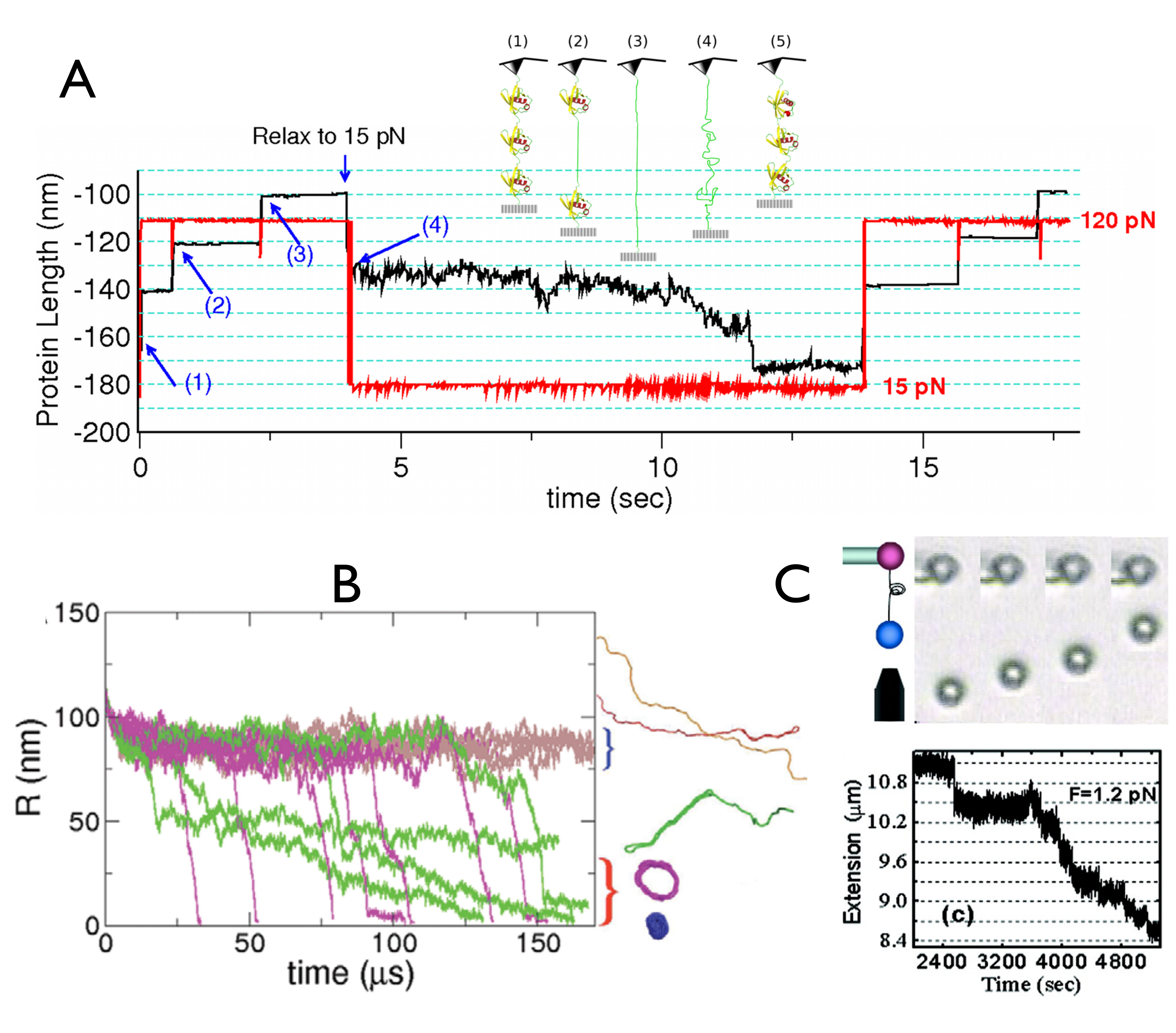}
 \caption{ }
\end{figure}

\end{document}